\documentclass[12pt,a4paper]{article}
\usepackage{graphicx}
\usepackage{sidecap}
\usepackage{caption}
\usepackage[font=small,labelfont=bf]{caption}

\begin{document}
\begin{center}
{\bf \Large Two photon exchange contribution \\ to small angle $\mu p$ and $ep$ elastic  scattering}\\
\vspace{0.3cm}
 M.G. Ryskin \\
 \vspace{0.2cm}
 Petersburg Nuclear Physics Institute, NRC Kurchatov Institute, Gatchina, Leningrad district,
 188300, Russia\\
 \vspace{0.5cm}
 {\bf Abstract:} We consider the role of the two-photon corrections to the small angle $\mu p$ and $ep$ elastic scattering  and the expected $\sigma(\mu^- p)/\sigma(\mu^+ p)$ ratio. 
\end{center}
\vspace{.5cm}
{\bf 1.} A new experiment~\cite{AV} on elastic electron-proton scattering at low momentum transfer squared $Q^2$
from 0.001 to 0.04 GeV$^2$ will be carried out by the PRES collaboration in the 720 MeV electron beam
of the Mainz Microtron MAMI. The goal of this experiment is to measure the proton charge radius with
sub-percent precision.

An analogous experiment is planned to be carried out at CERN in the 100 GeV muon beam by the
AMBER collaboration~\cite{Amb}. This will allow to compare with high precision the $ep$ and  $\mu p$ small angle
scattering studied under the same experimental conditions. Moreover, even the relativistic velocities of
the leptons (that is the $E/m_{lepton}$ ratios) will be close to each other. That means that these two
experiments will deal with almost the same classical currents.

The differential cross sections of elastic $ep$ and $\mu p$ scattering will be measured to the relative 0.1\% precision
in both experiments. This requires similar precision in the radiative corrections to the Born scattering
amplitude in the analysis of the experimental data. Note that the momentum transfer $Q^2$ will be
measured in these experiments via the energy of the recoiled protons detected in the Hydrogen Time
Projection Chamber. An important advantage of this method is strong suppression of the radiative
corrections cause by the lepton radiation (diagrams v2, r1,r2 in Fig.1)~\cite{Fad}. In this case,the main
contribution to the total radiative correction comes from the vacuum polarization and from the two
photon exchange diagrams v1, v4, v5. The vacuum polarization correction is of order of 1\% in the
considered $Q^2$ range, and it can be calculated with very high precision (~ 0.001\%). The situation with the two photon exchange correction is less certain. In this note, we present an analysis of the status of
theoretical calculations of this correction.
\vspace{3.0cm}
% Figure1-----------------------------------
\begin{figure}[ht]
%\vspace{-3.5cm}
\center{\includegraphics[width=100mm]{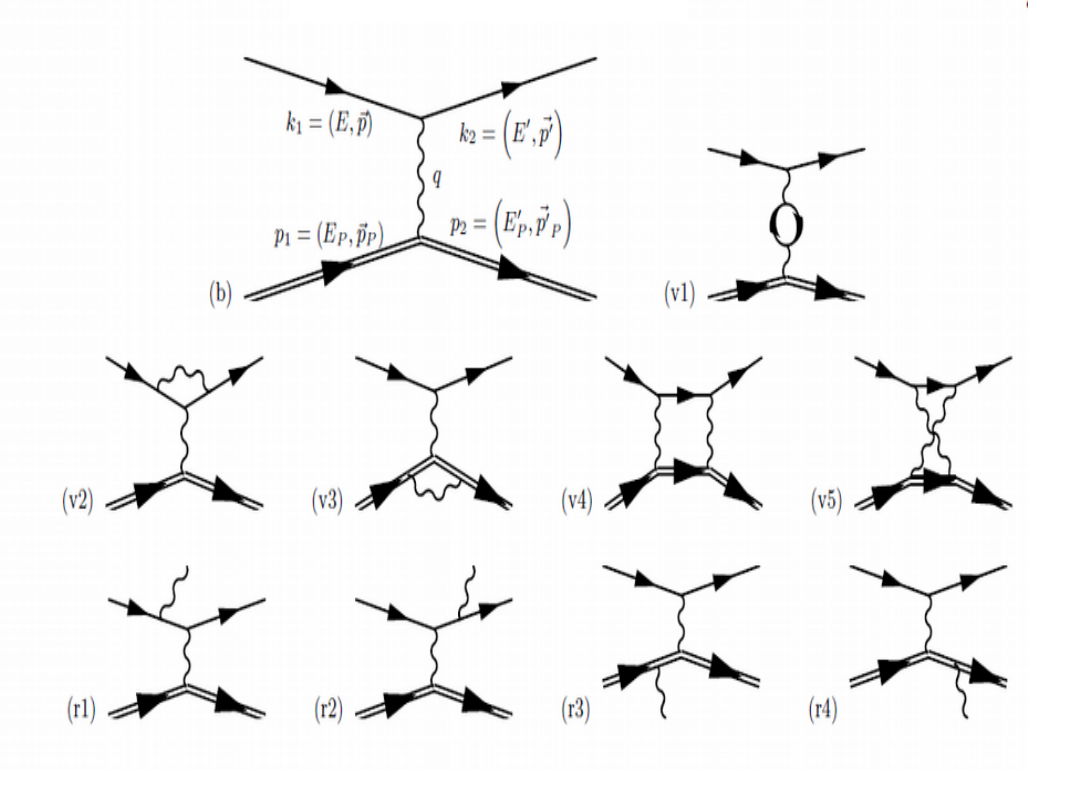}}

\caption{\sf The diagrams for the lepton-proton amplitude up to $\alpha^2$ order.}.
\end{figure}
% Figure1-----------------------------------
\\
{\bf 2.} The two photon exchange order of $\alpha=1/137$ correction to the Born $\mu p$ ($ep$) scattering amplitude
was considered long ago~\cite{1}. For a recent discussion see e.g.\cite{2}. Here we are interested in the low momentum transfer $Q^2=|t|$ region.

The two-photon contribution is described by the ’box’ and ’cross box’ (Xbox) diagrams v4 and v5 (Fig.1).   
The diagram with only the proton intermediate state, $X=p$, between the two photons is called 'elastic', while a heavier intermediate states $X=\Delta, \, N^*,\, N+\pi,\, etc.$ give the 'inelastic' contribution. Both elastic and inelastic $\delta_{2\gamma}$ corrections vanish at $t\to 0$.\\
Let us define $\delta_{2\gamma}$ as
\vspace{-.5cm}

\begin{equation}
\frac{d\sigma}{dt}=\frac{d\sigma^{Born}}{dt}(1+\delta_{2\gamma})
\end{equation}
At small $|t|<<M^2_{proton}=M^2$ (but $|t|>>m^2_{lepton}$) the two-photon elastic contribution reads~\cite{1}:
\begin{equation}
\delta^{el}_{2\gamma}=\pi\alpha\frac{\sin(\theta/2)}{1+\sin(\theta/2)}\simeq\pi\alpha\frac{\sqrt{|t|}}{2E_{lab}}
\end{equation}
%(see eq(76) of 1911.10956 and \cite{L})\\
The next term is proportional to $|t|$ and is enhanced by a logarithm.
\begin{equation}
\label{e3}
\delta^{el}_{2\gamma}=\pi\alpha\frac{\sqrt{|t|}}{2E_{lab}}+\frac{\alpha}{\pi}\left[\frac{|t|}{ME_{lab}}L(L+1)+O(|t|/M^2)\right]\  ,
\end{equation}

where $L=\ln(\sqrt{|t|}/2E_{lab})$ (see e.g.~\cite{14} eq.(50)).
\\

For the case of a large muon energy $E=E_\mu=100$ GeV and $t=-0.08$ GeV$^2$ the first term gives $\delta^{el}_{2\gamma}\simeq 0.3\cdot 10^{-4}$ and the second term is $\sim 1\cdot 10^{-4}$. This is negligible. However, for the lower electron energy of 720 MeV this two-photon 'elastic' correction is larger, reaching  $\delta^{el}_{2\gamma}\sim 0.004$ at $|t|=0.06$ GeV$^2$.
\\
Besides this, the inelastic correction may be not so small, in spite of the fact that at small $t\to 0$ $\delta^{inel}_{2\gamma}$ vanishes as
$\delta^{inel}_{2\gamma}=a|t|+b|t|\ln|t|$.
Using the dispersion relation and the leading logarithm approximation, we  get
for   $t=-0.08$ GeV$^2$  $\delta^{inel}_{2\gamma}$ = 0.0036 and  $\delta^{inel}_{2\gamma}$ = 0.0023  for the $ep$ and   $\mu p$ scattering, respectively.   
That  is  the  ratio  $\sigma(\mu^- p)$/$\sigma(\mu^+ p)$ 
is  expected to be  1 + 2 $\delta_{2\gamma}$ = 1.0046   at $t=-0.08$ $GeV^2$. 
In these calculations, the low $M_X$ region was approximated by the sum of the resonances (up to $M_X=1.8$ GeV) and by the Regge like parametrization with 
$\sigma(\gamma p)=(95+64/\sqrt{E_\gamma/1\mbox{GeV}})$ microbarn  
for $E_\gamma>1.3$ GeV~\cite{3}.
 This parametrization well describes the measured $\gamma p$
cross section,  as it can be seen in Fig.2.

% Figure2
 \vspace{-3cm}
\begin{figure}[h]
\center{\includegraphics[height=100mm]{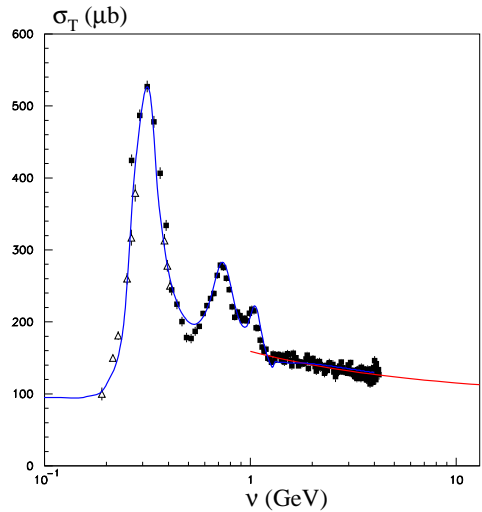}}
\caption{\sf The description of photon-proton cross section. The data are taken from~\cite{dat}}.
\end{figure}
 \vspace{-0.5cm}
The resulting corrections $\delta_{2\gamma}$ and the expected ratio  $R=\sigma(\mu^-p)/\sigma(\mu^+p)$ are shown in Fig.3 for the muons and in Fig.4 for the electrons.
Recall that the value of $\delta^{inel}_{2\gamma}$ was calculated here using the dispersion relation and the leading logarithmic approximation assuming  the zero subtraction constant at $E=0$. 
Clearly, there should be some corrections due to the non-logarithmic (NLO) terms and due to possible improvements in  precision of  high $M_X$  behaviour of the $\gamma p$ amplitude. But these corrections are not expected to be too large.

% Figure3
\begin{figure}[h]
 \vspace{-5cm}
\begin{center}
\hspace{-0.5cm}
\includegraphics[scale=0.33]{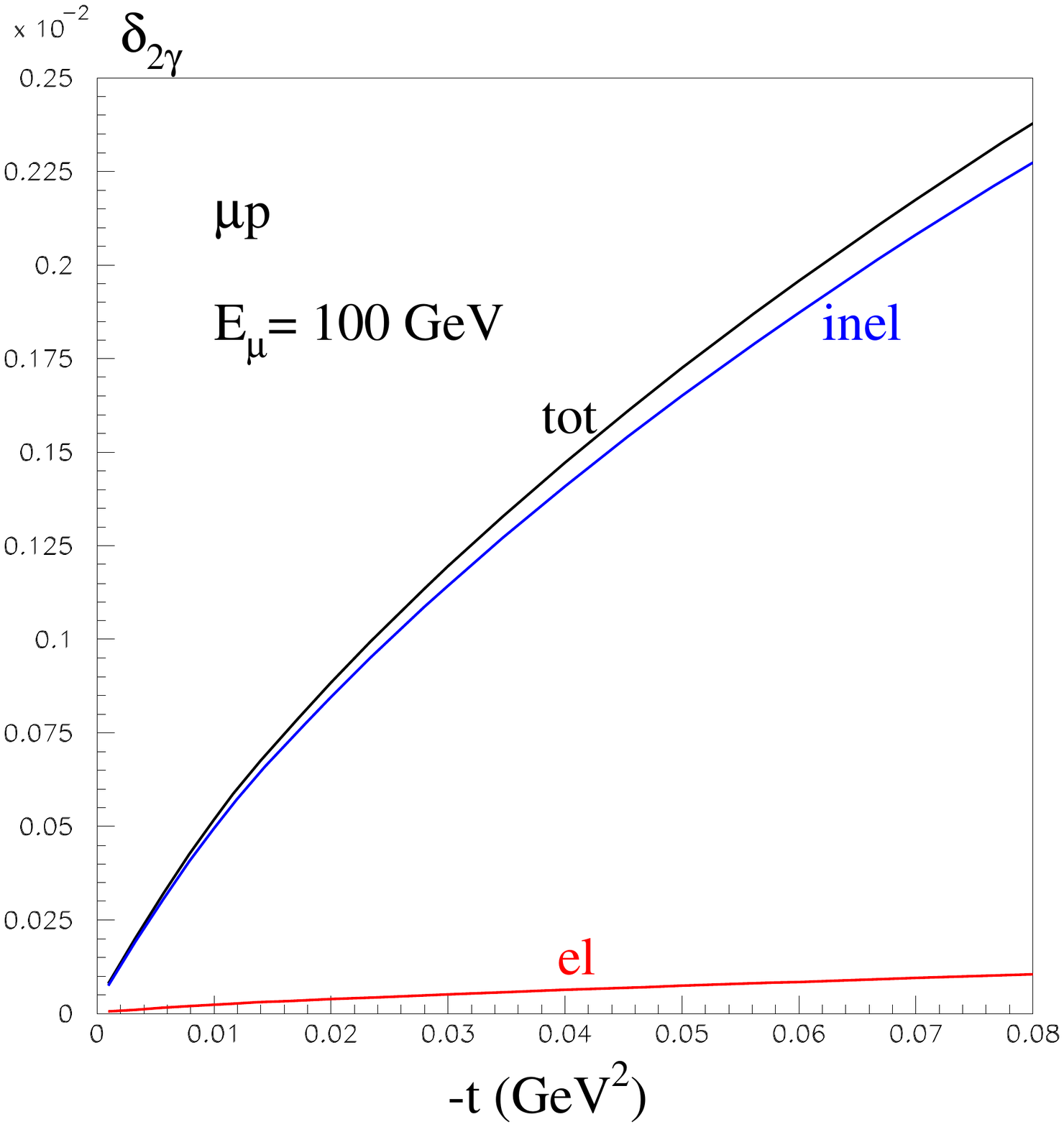}
\hspace{-0.5cm}
\includegraphics[scale=0.33]{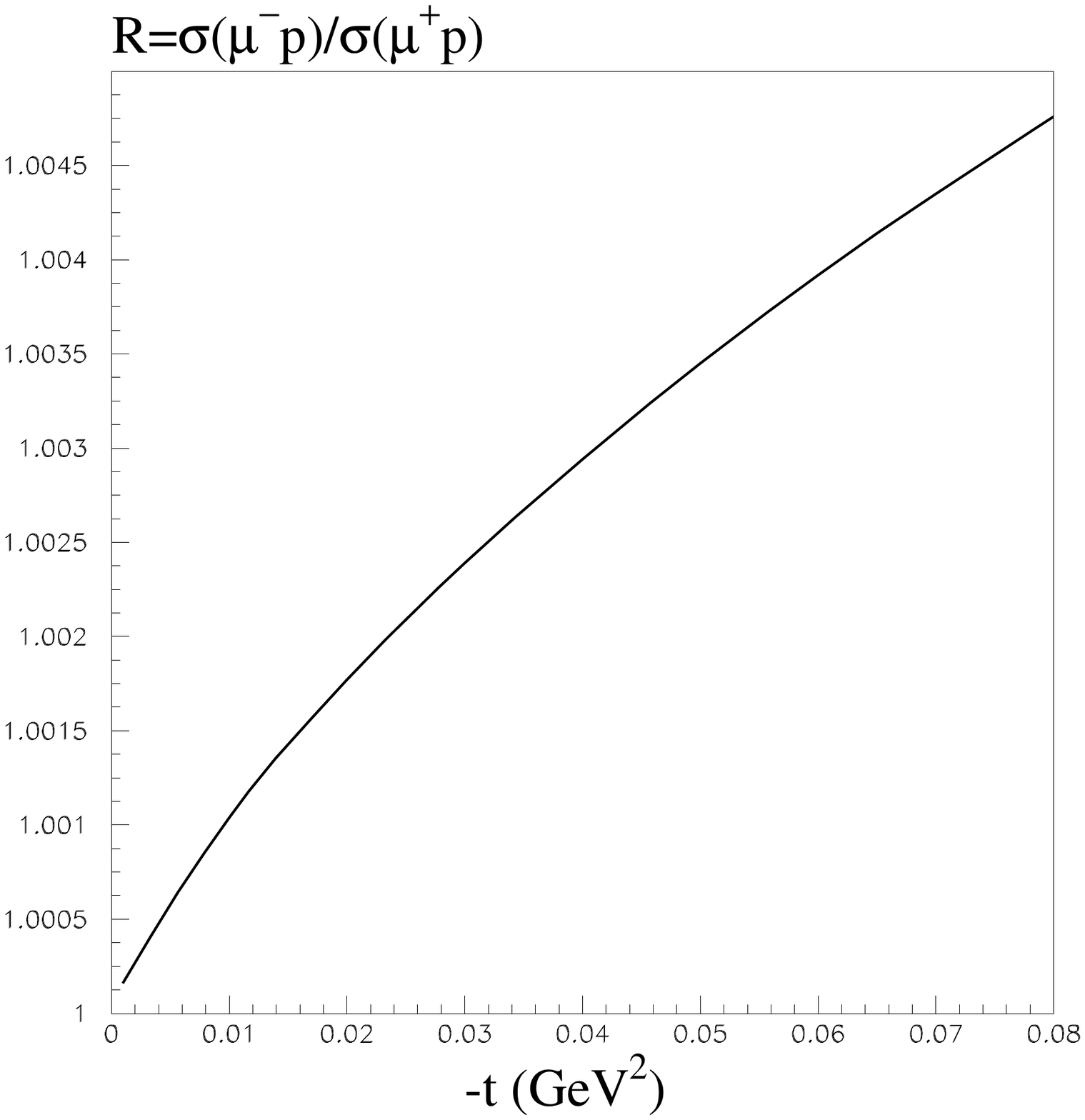}
%\hspace{-4.5cm}
%\includegraphics[width=8.5cm]{rl3}
\vspace{-0.2cm}
\caption{\sf Two-photon correction to the muon-proton cross section at $E_\mu=100$ GeV (left) and the ratio of $\mu^-p$ to $\mu^+p$ cross sections (right).}.
\end{center}
\end{figure}

% Figure4
\begin{figure}[tbh]
\vspace{-6cm}
\begin{center}
\hspace{-0.5cm}
\includegraphics[scale=0.33]{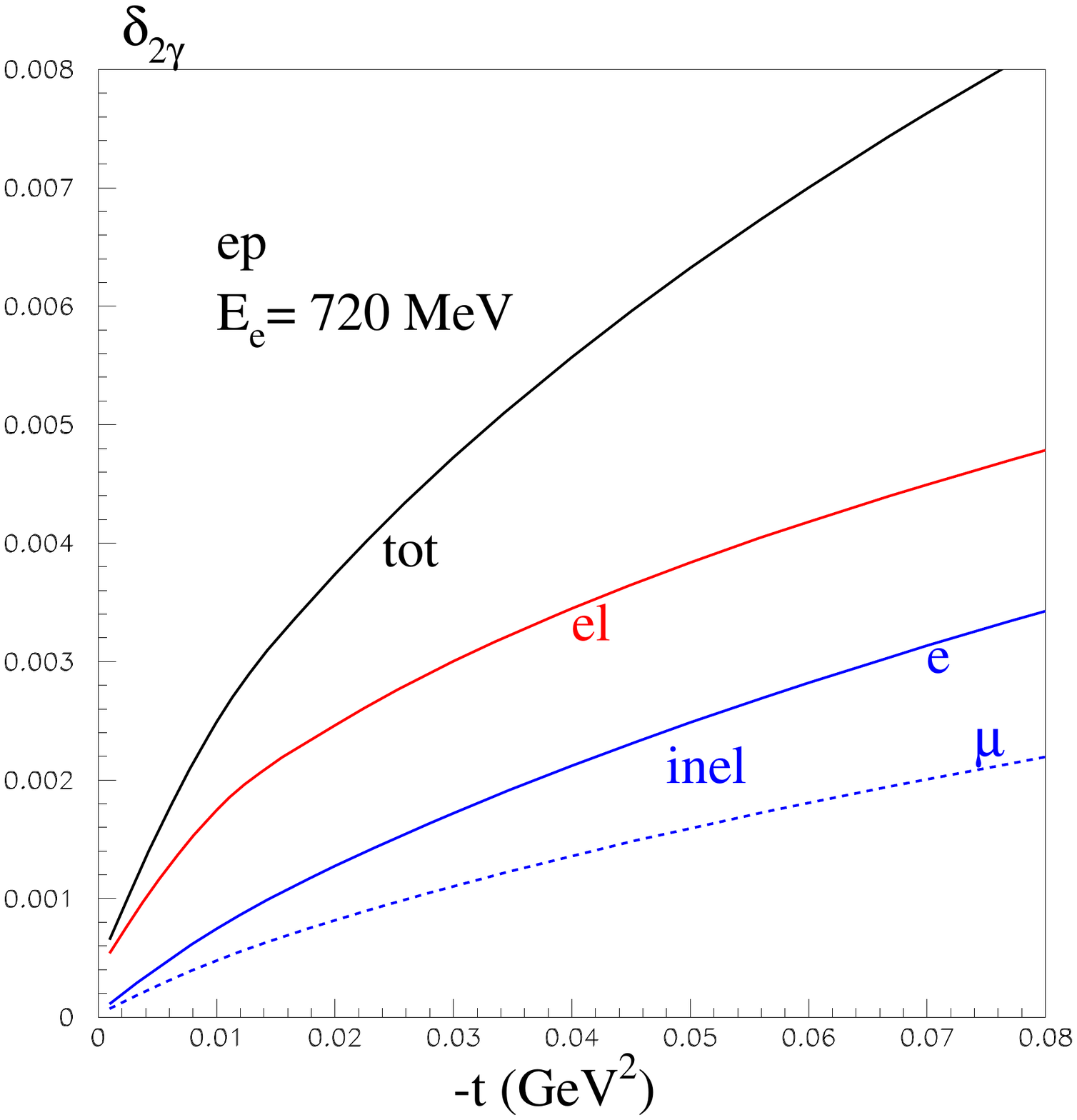}
\hspace{-0.5cm}
\includegraphics[scale=0.33]{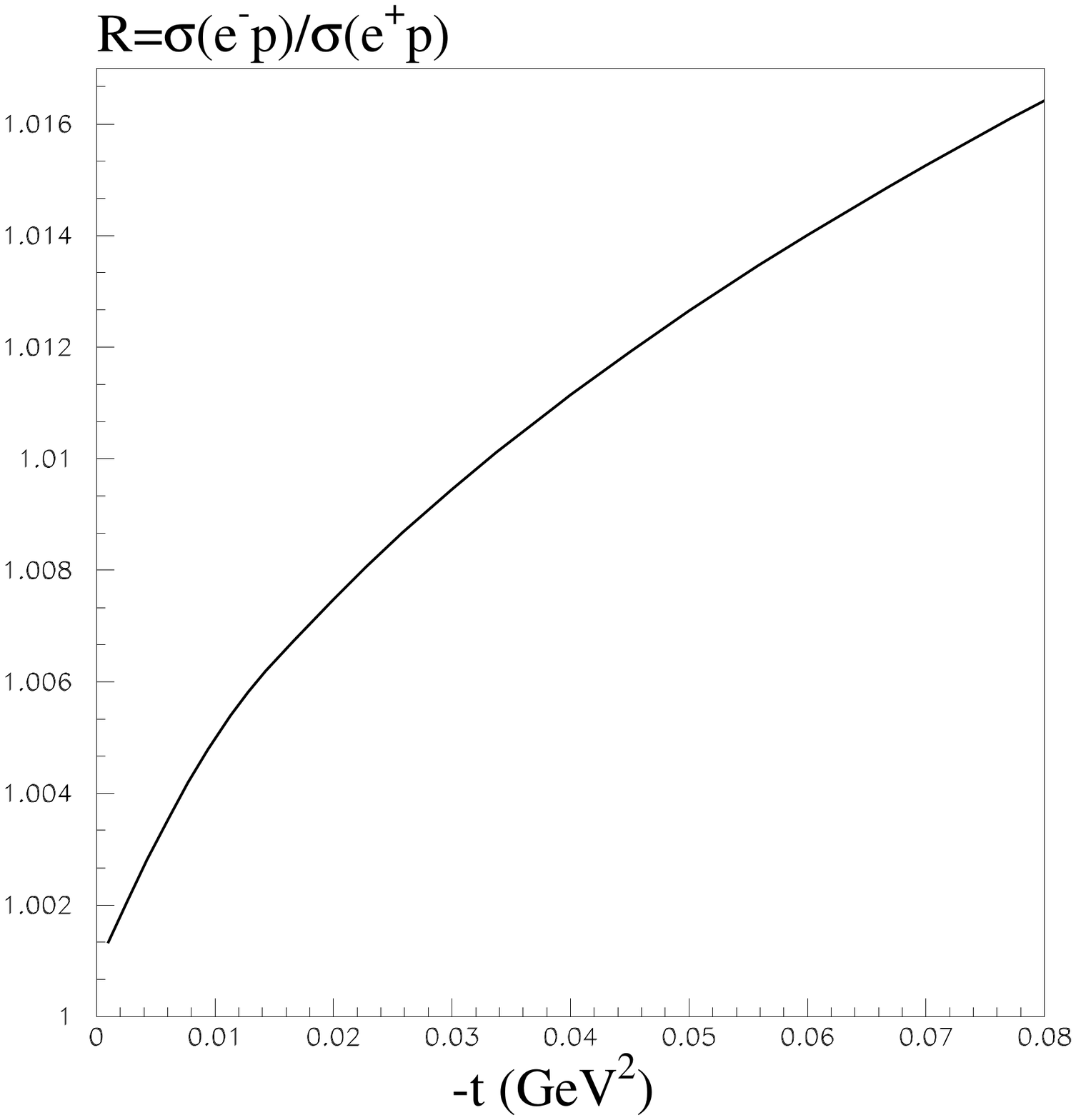}
\caption{\sf Two-photon correction to the electron-proton cross section at $E_e=720$ MeV (left) and the ratio of $e^-p$ to $e^+p$ cross sections (right). For comparison, the inelastic $\delta_{2\gamma}^{inel}$ correction for the muon-proton scattering with $E_\mu=100$ GeV is shown in the left panel by the dashed curve.}.
\end{center}
\end{figure}

\end{document}